\documentclass{optica-article}

\journal{opticajournal} 

\articletype{Research Article}

\usepackage[utf8]{inputenc}
\usepackage[english]{babel}
\usepackage[T1]{fontenc}
\usepackage{soul}
\usepackage[normalem]{ulem}
\usepackage{tikz}
\usepackage{lipsum}

\usepackage{graphicx,amsmath,color,bm,rotating,dsfont,stackrel,ulem,braket}
\usepackage[hyphenbreaks]{breakurl}
\usepackage{hyperref}

\newcommand{\ha}{\hat{a}}
\newcommand{\had}{\hat{a}^\dagger\vphantom{a}}
\newcommand{\hb}{\hat{b}}

\newcommand{\Tr}{\mathop{\mathrm{Tr}} \nolimits}

\newcommand{\vac}{\ket{\mathrm{vac}}}

\newcommand{\sugg}[1]{{#1}}

\begin{document}
	
	\title{
		Correlations 
		for \sugg{subsets of particles in} symmetric states: what photons are doing within a beam of light \sugg{when the rest are ignored}}
	
	\author{Aaron Z. Goldberg,\authormark{1,2,*}}
	
	\address{\authormark{1}National Research Council of Canada, 100 Sussex Drive, Ottawa, Ontario K1N 5A2, Canada\\
		\authormark{2}Department of Physics, University of Ottawa, Advanced Research Complex, 25 Templeton Street, Ottawa, Ontario K1N 6N5, Canada}
	
	\email{\authormark{*}aaron.goldberg@nrc-cnrc.gc.ca} 
	
	
	\begin{abstract*} 
		Given a state of light, how do its properties change when only some of the constituent photons are observed \sugg{and the rest are neglected (traced out)}? By developing formulae for mode-agnostic removal of photons from a beam,
		we show how the expectation value of any operator changes when only $q$ photons are inspected from a beam, ignoring the rest. We use this to reexpress expectation values of operators in terms of the state obtained by randomly selecting $q$ photons. Remarkably, this only equals the true expectation value for a unique value of $q$: expressing the operator as a monomial in normally ordered form, $q$ must be equal to the number of photons annihilated by the operator. 
		%
		A useful corollary is that the coefficients of any $q$-photon state chosen at random from an arbitrary state are exactly the $q$th order correlations of the original state; one can inspect the intensity moments to learn what any random photon will be doing and, conversely, one need only look at the $n$-photon subspace to discern what all of the $n$th order correlation functions are.
		The astute reader will be pleased to find no surprises here, only mathematical justification for intuition
		.
		Our results hold for any completely symmetric state of any type of particle with any combination of numbers of particles and can be used wherever bosonic correlations are found.
	\end{abstract*}
	
	\section{Introduction}
	Photodetection and photon statistics have been the backbone of quantum optics since its inception \cite{Glauber1963,Glauber1963quantumtheorycoherence,Glauber1963correlations,Sudarshan1963,MehtaSudarshan1965,TitulaerGlauber1965,CahillGlauber1969,AgarwalWolf1970,CarmichaelWalls1976,Loudon1976,Kimbleetal1977,Leuchs1986,Grangieretal1986,Rempeetal1990}. 
	For both single- and multi-mode states, correlation functions are routinely used to characterize the quantum or nonclassical nature of the field \cite{Mandel1979,Mandel1986,AgarwalTara1991,Klyshko1996,Lee1998,RichterVogel2002,ShchukinVogel2005,Achillesetal2006,Zavattaetal2007,Miranowiczetal2010,Ryletal2015,SahotaQuesada2015,Hartmannetal2015,Perinaetal2017,Sudhiretal2017,KorolkovaLeuchs2019,Malpanietal2019,Madsenetal2022}, fundamentally requiring simultaneous detection of multiple photons. 
	
	Photons, like all bosons, are totally symmetric under particle exchange, so a photodetector should be agnostic as to which photons it perceives. Sill, a detector seldom registers all of the photons in a beam, especially since beams tend to possess photon-number uncertainty, making the properties of subsets of photons from a beam crucial to its characterization. How do relevant quantities such as photon correlations change when some number of photons are removed from a beam [answered in Eq.~\eqref{eq:correlations Tr1 rhoN}] or when only a certain number of photons from the beam are detected [Eq.~\eqref{eq:correlation from q photons}]? How many photons must be registered to learn about particular properties of the beam [Eqs.~\eqref{eq:correlation from correct number of photons}, \eqref{eq:correlation from correct number of photons v2}]? Do the measured properties differ when different numbers of photons are detected [Eq.~\eqref{eq:rho random q}]? All of these questions have intuitive answers that follow directly from the framework we detail here.
	
	A familiar starting point for our investigation is the correspondence between the Poincar\'e and Bloch spheres. Classically, a quasi-monochromatic beam of light is represented by the ``Stokes vector'' pointing somewhere within the unit sphere named after Poincar\'e, while a qubit state such as a single photon's polarization state is represented by a ``Bloch vector'' lying within Bloch's sphere. Intuition says that a single photon chosen at random from a classical beam of light should have its Bloch vector correspond to the Stokes vector of the entire beam, and this is indeed the case \cite{Goldberg2022}. What happens when more than one photon is selected, especially in the case of quantum polarization where there is more polarization information encoded beyond the Stokes vector \cite{Klyshko1997,Luis2016,Goldbergetal2021polarization}? How does this intuition generalize when there are more than two polarization modes available to the photons, such as spatial or spectral modes? Not only are all of these questions answered here, they also do not depend on whether or not the beam of light has a determinate number of photons to begin with.
	
	The polarization case also exemplifies the crucial role of correlations in quantum optics: the Stokes vector and the beam's intensity are in one-to-one correspondence with the set of first-order correlations between the two polarization modes. These are the first-order degrees of coherence $g^{(1)}$ that are relevant for any pair of modes and are fundamental to experiments as basic as Young's double slit \cite{Young1801}. Since the correspondence implies that single photons inspected from a beam are sufficient for revealing $g^{(1)}$ functions, how many photons are required for measuring an arbitrary $g^{(n)}$ function? Could any $q$ photons be inspected to learn the $g^{(1)}$ functions? Our simple yet rigorous framework uniquely answers all of these questions in one fell swoop.
	
	Beyond intuition and mathematical completeness, the formulae we develop are useful for tasks such as quantifying quantumness in optical or other bosonic systems. Knowing how states behave after removing a certain number of photons is crucial to finding optimal quantum rotosensors \cite{ArnaudCerf2013,Baguetteetal2014,Giraudetal2015,BaguetteMartin2017} and to studies of photon loss in multimode systems such as boson sampling devices \cite{AaronsonBrod2016}. These tools can be added to the quantum optician's arsenal for a variety of applications.
	
	\sugg{Nomenclatural preliminaries are required. The removal of photons from a beam of light in different contexts give rise to the terms ``loss'' and ``photon subtraction'' that typically describe scenarios different from the present one. Photon subtraction, for example, is the result of acting on a state with an annihilation operator, which has a significant history \cite{Uedaetal1990,Daknaetal1997,Opatrnyetal2000,MizrahiDodonov2002,Wengeretal2004,Ourjoumtsevetal2006,Parigietal2007,Zavattaetal2008,Takahashietal2008,Gerritsetal2010,NavarreteBenlloch2012,Bartleyetal2013sub,FanZubairy2018,Stiesdaletal2021,Takaseetal2021,Melalkiaetal2022,Nunnetal2022} including experimental demonstrations that conditionally split off a single photon from a state at a weakly reflective beam splitter \cite{Wengeretal2004}. That action may indeed leave the overall state intensity unchanged, or even \textit{increased}, to appear as though no photons have been ``subtracted.'' Loss is the result of light coupling linearly with another mode that is later inaccessible; modelled using a beam-splitter transformation, it predictably reduces the intensity of input beams yet requires probabilistic combinations to handle situations in which definite numbers of photons have partial loss. Here, in contrast, we deal with beams that have an exact number of photons removed or that remove all photons until an exact number remain, which may be the result of ``discarding'' or ``ignoring'' or ``neglecting'' photons, or ``inspecting'' or ``focusing on'' a subset of photons, mathematically described by ``tracing out'' particles in a first-quantized picture. In the framework of states with a fixed number of photons, our scenario has actually been described as ``particle loss'' \cite{Nevenetal2018, Quintaetal2019} and as a ``fixed-loss model'' \cite{OszmaniecBrod2018}, terms we avoid due to their conflict with standard descriptions of optical loss. We thus re-emphasize: photons are indeed lost from our states and the resulting number of photons is indeed subtracted from the initial number, yet we refer to removal of photons from a beam in order to distance ourselves from the typical outcomes of loss or photon subtraction.}
	
	\section{Removing photons from a beam}
	We begin by considering a generic state ${\hat{\rho}}_N$ that contains exactly $N$ photons and asking what happens when one is removed. Considering each photon to potentially belong to one of $d$ modes, such a state is typically expanded in the Fock basis as
	\begin{equation}
		{\hat{\rho}}_N=\sum_{|\mathbf{m}|=|\mathbf{n}|=N}{\hat{\rho}}_{\mathbf{m}\mathbf{n}}\ket{\mathbf{m}}\bra{\mathbf{n}}.
	\end{equation} The basis is comprised of states
	\begin{equation}
		\ket{\mathbf{n}}=\frac{\ha_1^{\dagger n_1}}{\sqrt{n_1!}}\frac{\ha_2^{\dagger n_2}}{\sqrt{n_2!}}\cdots \frac{\ha_d^{\dagger n_d}}{\sqrt{n_d!}} \vac
	\end{equation} with $|\mathbf{n}|\equiv \sum_{i=1}^d n_i$, where the creation operators satisfy the usual bosonic commutation relations $[\ha_i,\ha^\dagger_j]=\delta_{ij}$. To discuss removal of a photon, we must rewrite each basis state in a first-quantized picture, with a completely symmetric sum over having $n_i$ of the photons being in state $\ket{i}$:
	\begin{equation}
		\ket{\mathbf{n}}=\sqrt{\frac{1}{\binom{N}{\mathbf{n}}}}\sum_{\mathrm{permutations}}\ket{1}^{\otimes n_1} \otimes\ket{2}^{\otimes n_2}\otimes\cdots\otimes\ket{d}^{\otimes n_d}.
	\end{equation} All of the photons are in a symmetric state, so we fiducially choose to remove the first one. With this equivalence, we can immediately see that projecting the first photon onto some state $\ket{i}$ just lowers the index $n_i$ by one in the second-quantized notation:
	\begin{equation}
		(\langle i|\otimes\mathds{1}^{\otimes N-1})\ket{\mathbf{n}}=\sqrt{\frac{n_i}{N}}\ket{n_1,n_2,\cdots,n_{i-1},n_i-1,n_{i+1},\cdots,n_d} =\frac{\ha_i}{\sqrt{N}}\ket{\mathbf{n}}.
	\end{equation} By linearity, any pure state gets projected to $(\langle i|\otimes\mathds{1}^{\otimes N-1})\ket{\psi}=\frac{\ha_i}{\sqrt{N}}\ket{\psi}$. It immediately follows that tracing out the first photon from a state leads to a convex combination of such operations and, since the first photon is equivalent to the rest, we find the generic relation
	\begin{equation}
		\Tr_{1}({\hat{\rho}}_N)=\frac{1}{N}\sum_{i=1}^d \ha_i{\hat{\rho}}_N \ha_i^\dagger.
	\end{equation} Removing another photon repeats the procedure with $N\to N-1$, so we employ the notation $\Tr_{k}({\hat{\rho}})$ to denote the removal of $k$ photons from a state ${\hat{\rho}}$ (i.e., a state $\Tr_{N-q}({\hat{\rho}}_N)$ has $q$ photons remaining \sugg{and is the result of inspecting only $q$ photons from the beam}).
	
	Going beyond a fixed $N$, the above relationships can be supplanted by 
	$\langle i\ket{\mathbf{n}}=\ha_i\frac{1}{\sqrt{\hat{N}}}\ket{\mathbf{n}}$, where $\hat{N}=\sum_{i=1}^d\ha_i^\dagger \ha_i$ is the total photon-number operator. For an arbitrary state, it then follows that
	\begin{equation}
		\Tr_{1}({\hat{\rho}})=\sum_{i=1}^d \ha_i\frac{1}{\sqrt{\hat{N}}}{\hat{\rho}} \frac{1}{\sqrt{\hat{N}}} \ha_i^\dagger=\sum_{i=1}^d \frac{1}{\sqrt{\hat{N}+1}}\ha_i{\hat{\rho}}  \ha_i^\dagger \frac{1}{\sqrt{\hat{N}+1}},
		\label{eq:Tr1 general}
	\end{equation} which offers a unique way of removing photons even from superposition states that carry terms of the form $\ket{\mathbf{m}}\bra{\mathbf{n}}$ with $|\mathbf{m}|\neq|\mathbf{n}|$. This quantity is unchanged via mode transformations and repartitionings of Hilbert space $\ha_i\to\sum_{j=1}^d U_{ij}\ha_j$ for unitary matrices $U$, as expected for a mode-agnostic description of removing photons.
	
	\sugg{Equation~\eqref{eq:Tr1 general} properly describes how to treat a beam of light that has exactly one photon removed; beyond providing a mathematical language for discussing such situations, do the latter arise physically? One can argue that they occur pervasively in nature: when imaging a beam of light with a camera or one's eye, not all of the photons are always recorded, even if no photons are lost on the way and the detector is perfectly efficient. This occurs, for example, in coincidence detection, when the one records events in which two photons from a generic beam of light together impinge on any detector. In that case, one is inspecting a beam that effectively had photons removed according to our formalism. One could consider \textit{simulating} Eq.~\eqref{eq:Tr1 general} using mode-agnostic ``photon subtraction'' \cite{Melalkiaetal2022} to enact each operation $\rho\to \ha_i\rho\ha_i^\dagger$ and then normalizing the resultant state using a complicated photon-number-dependent operation that acts on all states as $|\psi\rangle\to \frac{1}{\sqrt{\hat{N}+1}}|\psi\rangle$, but this is physically a bit forced and is better considered using the simple photon-subset physical scenario just described.}
	
	The above and ensuing formulae can all be applied to continuous modes by replacing the sums with integrals and the Kronecker-delta commutation relations with Dirac-delta ones. One can also imagine applying the same formula Eq.~\eqref{eq:Tr1 general} to fermionic or anyonic states, but one loses some of the richness upon noticing that a single fermionic mode can only be annihilated once before vanishing.
	
	\section{How correlations change with photon removal}
	Given an observable quantity, we expect a reasonable relationship between the quantity measured for the entire state and the quantity measured for the state with one or more photons removed. We start with a simple first-order correlation within one mode or between two modes, $\langle \ha_k^\dagger \ha_l\rangle$, and ask how the expectation value changes when taken with respect to ${\hat{\rho}}_N$ versus $\Tr_{1}({\hat{\rho}}_N)$. By linearity, these expressions will tell us how any generator of the group SU$(d)$ changes when one $d$-level particle is removed from an $N$-particle symmetric state.
	
	It is easy to compute, using the bosonic commutation relations,
	\begin{equation}\begin{aligned}
			\langle\ha_k^\dagger\ha_{l}\rangle_{\Tr_{1}({\hat{\rho}}_N)}&=\Tr(\ha_k^\dagger\ha_{l}\frac{1}{N}\sum_{i=1}^d \ha_i{\hat{\rho}}_N \ha_i^\dagger)
			\\&
			=\frac{1}{N}\sum_{i=1}^d\Tr[\ha_k^\dagger(\ha_{l}\ha_i^\dagger-\delta_{il}) \ha_i{\hat{\rho}}_N ]\\
			&=\frac{1}{N}\Tr[\ha_k^\dagger\ha_{l}(\hat{N}-1) {\hat{\rho}}_N ]
			=\frac{N-1}{N}\langle\ha_k^\dagger\ha_{l}\rangle_{{\hat{\rho}}_N}.
			\label{eq:tracing one photon first order correlation}
	\end{aligned}\end{equation} Since this can be repeated for removing more photons, this means that the \textit{ratios} between all of the first-order correlations for a state ${\hat{\rho}}_N$ are unchanged by removing any number of photons. As well, since the first-order correlations set the value of $\langle\hat{N}\rangle$, the normalization of all of the first-order correlations can be known directly from the total number of photons measured. For the example of the Stokes vector, we see that the classical polarization properties of a beam with a fixed number of photons are unchanged when only a subset of the photons are measured.
	
	There are two immediate directions to generalize this calculation: to more general operators than first-order correlations and to more general states without fixed photon numbers.
	
	A generic operator can always be expressed in normally ordered form \cite{Mikhailov1983,Blasiaketal2003} with components
	\begin{equation}
		\hat{O}_{\mathbf{k}\mathbf{l}}=\ha_1^{\dagger k_1}\ha_2^{\dagger k_2}\cdots \ha_d^{\dagger k_d}\ha_1^{l_1}\ha_2^{l_2}\cdots \ha_d^{l_d}.
	\end{equation} Note that each $\ha_i$ acts on mode $i$ such that many of these operators in the product commute with each other. Then, the same calculation as Eq.~\eqref{eq:tracing one photon first order correlation} but using
	the more general relationship $[\ha_i^{l_i},\ha_i^\dagger]=l_i \ha_i^{l_i-1}$
	dictates that
	\begin{equation}
		\langle \hat{O}_{\mathbf{k}\mathbf{l}}\rangle_{\Tr_{1}({\hat{\rho}}_N)}=\frac{N-|\mathbf{l}|}{N}\langle \hat{O}_{\mathbf{k}\mathbf{l}}\rangle_{{\hat{\rho}}_N}.
		\label{eq:correlations Tr1 rhoN}
	\end{equation} For this expression to be nonzero, we also require $|\mathbf{k}|=|\mathbf{l}|$. This reinforces that one requires at least $|\mathbf{l}|$ photons to observe a correlation that has $|\mathbf{l}|$ annihilation or creation operators; after removing one photon at a time until $N=|\mathbf{l}|$, removing one more photon causes these correlations to all vanish from the resulting state.
	
	Convex combinations of the above results suffice to explain how the expectation values of all operators change upon removal of photons from states of the form $\sum_N p_N {\hat{\rho}}_N$ and of operators with $|\mathbf{k}|=|\mathbf{l}|$ for all states. What about expectation values of operators with $|\mathbf{k}|\neq|\mathbf{l}|$ for states with coherences between different photon-number sectors? We simply refrain from the substitution $\hat{N}\to N$ in the derivation of Eq.~\eqref{eq:tracing one photon first order correlation} to find many equivalent expressions:
	\begin{equation}
		\langle \hat{O}_{\mathbf{k}\mathbf{l}}\rangle_{\Tr_{1}({\hat{\rho}})}=
		\Tr(\hat{O}_{\mathbf{k}\mathbf{l}}\frac{\hat{N}-|\mathbf{l}|}{\sqrt{\hat{N}}}{\hat{\rho}} \frac{1}{\sqrt{\hat{N}}})=
		\Tr(\hat{O}_{\mathbf{k}\mathbf{l}}\frac{1}{\sqrt{\hat{N}}}{\hat{\rho}} \frac{\hat{N}-|\mathbf{k}|}{\sqrt{\hat{N}}})=
		\Tr(\hat{O}_{\mathbf{k}\mathbf{l}}\sqrt{\frac{\hat{N}-|\mathbf{l}|}{\hat{N}}}{\hat{\rho}} \sqrt{\frac{\hat{N}-|\mathbf{k}|}{\hat{N}}}).
	\end{equation} One can chooses which one of the formulae to apply, such as $\frac{\hat{N}-|\mathbf{l}|}{\sqrt{\hat{N}}}{\hat{\rho}} \frac{1}{\sqrt{\hat{N}}}$ to the state or $\frac{1}{\sqrt{\hat{N}}}\hat{O}_{\mathbf{k}\mathbf{l}}\frac{\hat{N}-|\mathbf{l}|}{\sqrt{\hat{N}}}$ to the operator, to see how the correlations change upon removal of a photon.
	To apply this formula, we take every term in ${\hat{\rho}}$ of the form $\ket{\mathbf{m}}\bra{\mathbf{n}}$ with $|\mathbf{m}|-|\mathbf{l}|=|\mathbf{n}|-|\mathbf{k}|$ that contributed to $\langle \hat{O}_{\mathbf{k}\mathbf{l}}\rangle$ and scale its contribution by $\frac{|\mathbf{m}|-|\mathbf{l}|}{\sqrt{|\mathbf{m}||\mathbf{n}|}}$.
	Note that the correlations do not simply get multiplied by the same factor, unlike the case for ${\hat{\rho}}_N$, such that the ratios of the correlations change upon removal of photons from a state and thereby the correlations change significantly when one only observes a subset of photons from a state.
	
	The correlations with $|\mathbf{k}|\neq|\mathbf{l}|$ slightly complicated all of the expressions. This seems to imply that measuring a quadrature operator such as $\ha_i+\ha_i^\dagger$ requires more careful calculation than an intensity correlation like $\ha_i^\dagger \ha_j+\ha_j^\dagger \ha_i$. However, it behooves one to recall that homodyne measurements that purport to measure quadrature operators on mode $i$ are actually measuring an intensity correlation between that mode and a field mode $j$, where the latter is taken to be in a coherent state with large amplitude. All routine measurements thus tend to fundamentally be made from operators that annihilate as many photons as they create, such that the simple expressions $\langle \hat{O}_{\mathbf{k}\mathbf{l}}\rangle_{\Tr_{1}({\hat{\rho}})}=\sum_N p_N\frac{N-|\mathbf{l}|}{N} \langle \hat{O}_{\mathbf{k}\mathbf{l}}\rangle_{{\hat{\rho}}_N}$ can typically be used.
	
	\sugg{A different sort of correlations were inspected in Refs.~\cite{Nevenetal2018,Quintaetal2019}, where states with fixed $N$ were inspected for their \textit{entanglement} properties when one or more photons are removed or ignored using the same mechanism as the present one. It is likely that various combinations of the correlations $\langle\hat{O}_{\mathbf{kl}}\rangle$ can be used to witness entanglement, but different combinations will be necessary as witnesses for different scenarios and so we leave that investigation for future study.}
	
	\section{Observing $q$ photons at a time}
\begin{figure}
		\centering
		\includegraphics[width=0.65\textwidth]{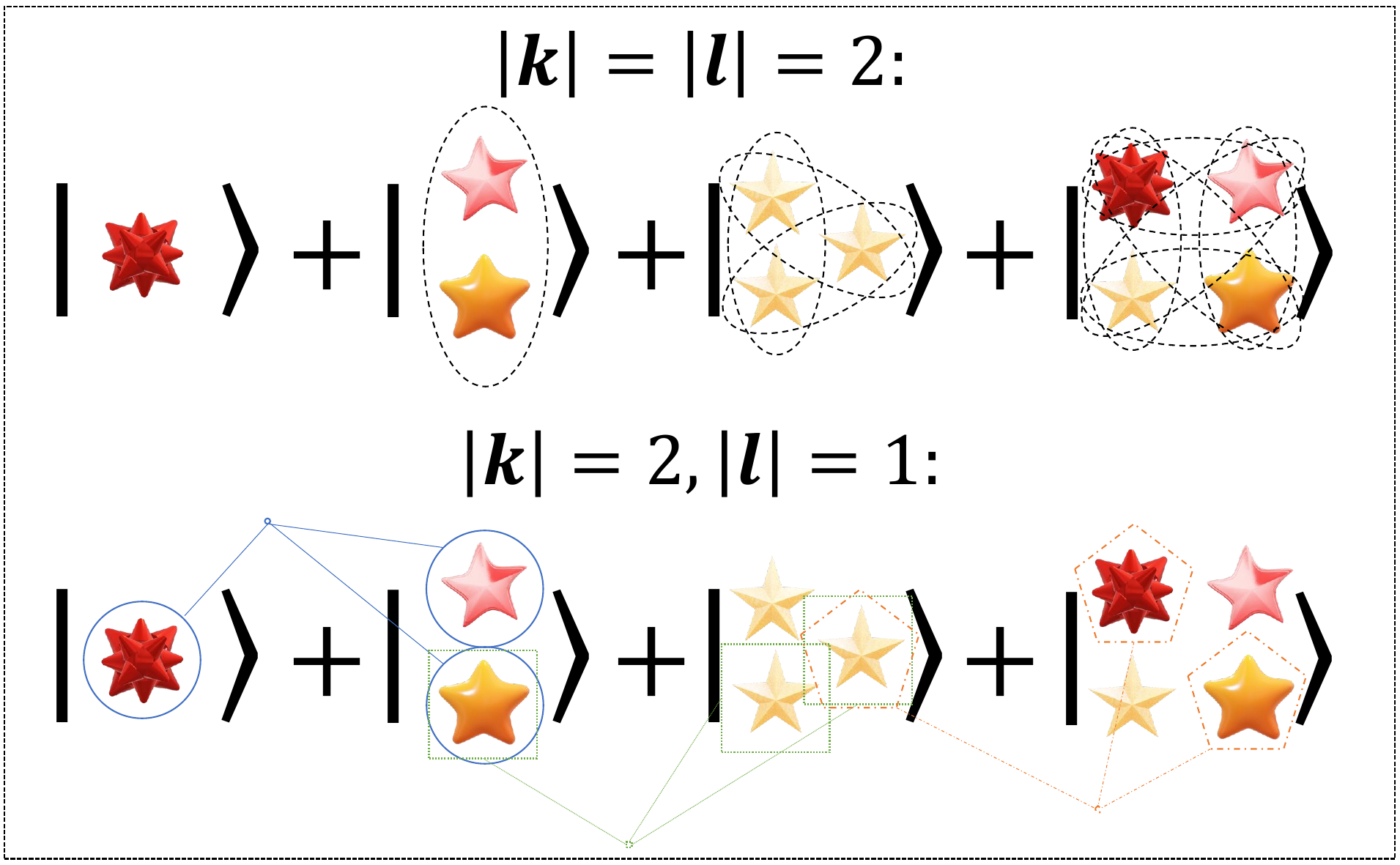}
		\caption{Schematizing which photons contribute to measuring which operators. A generic superposition state may be comprised of multiple photons (different kets) in multiple modes (different star icons). Inspecting two photons at a time (upper half of the figure) gives rise to all second-order correlations $\langle\hat{O}_{\mathbf{k}\mathbf{l}}\rangle$ with $|\mathbf{k}|=|\mathbf{l}|=2$, with all the relevant pairs of photons circled by the dashed ovals; the pairs must all come from the same ket. Inspecting another set of correlations for the same state, such as those with $|\mathbf{k}|=2$ and $|\mathbf{l}|=1$, requires comparing one photon from one ket and two photons from the ket with one more total photon (lower half of the figure); we include three representative combinations in blue circles, green squares, and orange pentagons. Generalizations to mixed states and superpositions with different amplitudes follow the intuition from this picture and are discussed in the text.}
		\label{fig:choosing photons}
	\end{figure}

	We argue here that there is only one possible integer number of photons $q$ that should be inspected at a time in order to measure any particular correlation $\langle \hat{O}_{\mathbf{k}\mathbf{l}}\rangle$ (see Fig.~\ref{fig:choosing photons}): \begin{equation}
		q=|\mathbf{l}|=|\mathbf{k}|.
	\end{equation} For operators with $|\mathbf{k}|\neq |\mathbf{l}|$, we simply assume they are homodyne-type operators whose reference modes should be reintroduced until $|\mathbf{k}|= |\mathbf{l}|$. We give an argument for what to do in cases where this is not possible at the end of this section (again, see Fig.~\ref{fig:choosing photons}).
	
	A generic state ${\hat{\rho}}$ can be rewritten in terms of components with fixed and non-fixed photon numbers:
	\begin{equation}
		{\hat{\rho}}=\sum_N p_N {\hat{\rho}}_N +\sum_{|\mathbf{m}|\neq |\mathbf{n}|}{\hat{\rho}}_{\mathbf{m}\mathbf{n}}\ket{\mathbf{m}}\bra{\mathbf{n}},
	\end{equation} where each ${\hat{\rho}}_N$ is a normalized density operator with $N$ photons and none of the terms in the latter sum contribute to expectation values of operators with $|\mathbf{k}|=|\mathbf{l}|$. Observing $q$ photons from a state ${\hat{\rho}}_N$ requires computing \begin{equation}
		{\hat{\varrho}}(q|{\hat{\rho}}_N)\equiv\Tr_{N-q}({\hat{\rho}}_N)
	\end{equation} by repeatedly using the simple formula Eq.~\eqref{eq:Tr1 general} from above. For the more general state ${\hat{\rho}}$, one must weigh each of these contributions by the probability that the photons were taken from that sector: \begin{equation}P(q\,\mathrm{from}\,N)\equiv\frac{1}{\mathcal{N}(q)} p_N\binom{N}{q},\end{equation} where $\mathcal{N}(q)=\sum_N p_N\binom{N}{q}$ is a normalization constant equal to the total number of expected $q$-photon events.
	
	It would be nice to express the expectation value of an observable in terms of the $q$ photons that are actually observed. The former should take the form of the value of the observable for $q$ photons that came from a given $N$-photon subspace $\langle \hat{O}_{\mathbf{k}\mathbf{l}}\rangle_{{\hat{\varrho}}(q|{\hat{\rho}}_N)}$, i.e. the expectation value with respect to the state $\Tr_{N-q}({\hat{\rho}}_N)$, multiplied by the probability that the photons came from that subspace, all scaled by the total number of times that $q$ photons are actually observed:
	\begin{equation}\begin{aligned}
			\langle \hat{O}_{\mathbf{k}\mathbf{l}}\rangle_{\hat{\rho}} \underset{?}{=}&\mathcal{N}(q)\sum_N P(q\,\mathrm{from}\,N) \langle \hat{O}_{\mathbf{k}\mathbf{l}}\rangle_{{\hat{\varrho}}(q|{\hat{\rho}}_N)}\\
			=&\sum_N p_N \binom{N}{q} \frac{q!}{N!}\frac{(N-|\mathbf{l}|)!}{(q-|\mathbf{l}|)!}\langle\hat{O}_{\mathbf{k}\mathbf{l}}\rangle_{{\hat{\rho}}_N}\\
			=&\sum_N p_N \binom{N-|\mathbf{l}|}{q-|\mathbf{l}|} \langle\hat{O}_{\mathbf{k}\mathbf{l}}\rangle_{{\hat{\rho}}_N}.
			\label{eq:correlation from q photons}
	\end{aligned}\end{equation} This equality is correct for arbitrary states if and only if $q=|\mathbf{l}|$. We thus learn that the expectation value of a correlation operator that annihilates $q$ photons from a variety of modes and creates $q$ photons among those modes can uniquely be reinterpreted as the the expectation value the operator takes for a randomly selected subset of $q$ photons from the state. We cement this relationship as
	\begin{equation}
		\langle \hat{O}_{\mathbf{k}\mathbf{l}}\rangle_{\hat{\rho}}=\mathcal{N}(|\mathbf{l}|)\sum_N P(|\mathbf{l}|\,\mathrm{from}\,N) \langle \hat{O}_{\mathbf{k}\mathbf{l}}\rangle_{{\hat{\varrho}}(|\mathbf{l}||{\hat{\rho}}_N)}.
		\label{eq:correlation from correct number of photons}
	\end{equation}
	
	Immediate consequences include that all intensity moments must be measured one photon at a time. These include the Stokes vector, for example, and are simply scaled by the average number of photons $\mathcal{N}(1)=\langle \hat{N}\rangle_{\hat{\rho}}$. For a correlation function like $g^{(2)}$, two photons must be measured at a time and, similarly, for any $g^{(n)}$, $n$ photons must be inspected simultaneously.
	
	As a corollary, we can immediately conclude what the state of a density operator must be when $q$ photons are chosen at random from it. The coefficients of the density operator in this basis must be the values of the correlation functions given by the appropriate observables:
	\begin{equation}
		{\hat{\varrho}}(q|{\hat{\rho}})\equiv \sum_N P(q\,\mathrm{from}\,N) {\hat{\varrho}}(q|{\hat{\rho}}_N) \propto \sum_{|\mathbf{m}|=|\mathbf{n}|=q}\frac{\langle \hat{O}_{\mathbf{m}\mathbf{n}}\rangle_{{\hat{\rho}}}}{\sqrt{m_1!\cdots m_d!n_1!\cdots n_d!}}\ket{\mathbf{n}}\bra{\mathbf{m}}.
		\label{eq:rho random q}
	\end{equation} This is emphatically different from the projection of a state ${\hat{\rho}}$ onto the $q$-photon subspace and allows us to rewrite Eq.~\eqref{eq:correlation from correct number of photons} as \begin{equation}\langle \hat{O}_{\mathbf{k}\mathbf{l}}\rangle_{\hat{\rho}}=\mathcal{N}(|\mathbf{l}|)\langle \hat{O}_{\mathbf{k}\mathbf{l}}\rangle_{{\hat{\varrho}}(|\mathbf{l}||{\hat{\rho}})}.
		\label{eq:correlation from correct number of photons v2}
	\end{equation} The factorial factors are necessary to cancel the values $\langle \mathbf{m}|\hat{O}_{\mathbf{m}\mathbf{n}}|\mathbf{n}\rangle$, while the proportionality constant is the inverse of $\mathcal{N}(q)=\sum_{|\mathbf{m}|=q}\frac{\langle \hat{O}_{\mathbf{m}\mathbf{m}}\rangle_{{\hat{\rho}}}}{m_1!\cdots m_d!}$.  We give a more formal proof of this relationship in Supplement 1, in case it is more appealing than our intuitive arguments for some readers.
	
	From this corollary we learn the intuitive equivalence: the expansion coefficients of a randomly chosen $q$ photons from a beam in the Fock space of the $d$ modes tell you what the state's $q$th-order correlations are and vice versa. Inspecting another number of photons at a time cannot tell you about other correlations and inspecting other correlations cannot tell you how a certain $q$ photons will behave. More forcibly, this means that, for example, should two photons be detected simultaneously among all of the possible modes, their statistics would say nothing about the intensity correlations in the beam from whence they originated.
	
	All of the above should suffice for operators with $|\mathbf{k}|\neq|\mathbf{l}|$ if appropriate reference modes are considered. We finally seek a construction that may work in the absence of such reference modes. There is no unambiguous meaning to tasks such as ``removing one photon from $\ket{2,2}\bra{1,1}$'' but we can apply our $\Tr_1$ formula and see what ensues. Performing the same calculations as above, 
	we find the unique equivalence
	\begin{equation}
		\langle \hat{O}_{\mathbf{k}\mathbf{l}}\rangle_{\hat{\rho}}=\sum_{\mathbf{m}\mathbf{n}}\sqrt{{\hat{\rho}}_{\mathbf{m}\mathbf{n}}\binom{|\mathbf{m}|}{|\mathbf{l}|}}\sqrt{{\hat{\rho}}_{\mathbf{m}\mathbf{n}}\binom{|\mathbf{n}|}{|\mathbf{k}|}}
		\langle\hat{O}_{\mathbf{k}\mathbf{l}}\rangle_{\Tr_{|\mathbf{m}|-|\mathbf{l}|}(\ket{\mathbf{m}}\bra{\mathbf{n}})}
		.
	\end{equation} The expectation value for the overall state is the same as the the expectation value in the reduced state that has $|\mathbf{k}|$ photons remaining in the bra and $|\mathbf{l}|$ photons remaining in the ket, weighed by a sort-of probability amplitude that the photons came from the $|\mathbf{m}|$- and $|\mathbf{n}|$-photon subspaces. No other terms can contribute to this expectation value (notice that $|\mathbf{n}|-|\mathbf{k}|$ must equal $|\mathbf{m}|-|\mathbf{l}|$) and this is the unique weight distribution that allows for a reinterpretation of expectation values for all states. We can also use this to construct some sector of a density operator that has ``$q_1$ photons in its bras and $q_2$ photons in its kets'' whose coefficients would be the corresponding correlations $\langle \hat{O}_{\mathbf{k}\mathbf{l}}\rangle_{\hat{\rho}}$, should we be willing to slightly abuse notation. Any other construction, however, would be disingenuous. 
	
	\section{Agreement with intuition from photodetection}
	The origins of the correlation functions $\langle \ha_i^{\dagger l_i}\ha_i^{l_i}\rangle$ stem from a photodetection model where $l_i$ photons are absorbed by the detector and thereby annihilated from mode $i$. All of the remaining correlation functions to the same order can be obtained via a similar argument among a variety of modes $\ha_i\to\sum_{j=1}^d U_{ij}\ha_j$, so the most basic argument of standard photodetection is that $|\mathbf{l}|$ photons must be absorbed by a detector to measure a correlation of order $|\mathbf{l}|$. In this work, we have shown a separate, equivalent, mathematically rigorous method for motivating the same physical picture: the state of a beam of light that would be obtained by randomly selecting $|\mathbf{l}|$ photons to be sent to a detector is exactly the state that exclusively contains the information about the $|\mathbf{l}|$th order correlations. Photodetection removes $|\mathbf{l}|$ photons to inspect them; our work shows that the $|\mathbf{l}|$ photons exactly carry the appropriate correlation information. As heralded in the abstract, the astute quantum optician will be pleased to find no surprises here, only justification for intuition.
	
	\section{Incorporating and comparing to standard optical loss}
	How does one deal with the situation in which $q$ photons are observed simultaneously, but some number of photons $r$ were lost prior to the detection? Should one use the state ${\hat{\varrho}}(q+r|{\hat{\rho}})$ or ${\hat{\varrho}}(q|{\hat{\rho}})$ for predicting what values the correlations will take?
	
	There is actually no ambiguity here: one simply uses the state ${\hat{\rho}}$ that arises from the original state having lost $r$ photons and inputs it into the formula for ${\hat{\varrho}}(q|{\hat{\rho}})$. If one uses the state before it lost the $r$ photons, one must be able to have access to the loss modes. Then one would simply increase $d$ and detect $q$ photons among all of the modes; given that loss modes tend to be lost, this method will seldom work and one should instead treat the state as having lost $r$ photons prior to computing ${\hat{\varrho}}(q|{\hat{\rho}})$.
	
	The other question is how this work relates to standard loss channels, which enact the input-output relations
	\begin{equation}
		\ha_i\underset{\eta_i}{\to}\sqrt{\eta_i}\ha_i+\sqrt{1-\eta_i^2}\hb_i
	\end{equation} and then trace over the ``vacuum modes'' annihilated by $\hb_i$ that begin in their vacuum states. We show in Supplement 2 that having equal loss $\eta_i=\eta$ in all $d$ modes is equivalent to a state undergoing the transformation
	\begin{equation}\begin{aligned}
			\hat{\rho}_N\underset{\eta}{\to}&
			\sum_k\binom{N}{k}\left(1-\eta\right)^{k} \eta^{N-k}
			\Tr_{N-k}(
			\hat{\rho}_N);\\
			\hat{\rho}\underset{\eta}{\to}
			&\sum_k\frac{\left(\frac{1-\eta}{\eta}\right)^{k}}{k!} 
			\underbrace{\Tr_1(\sqrt{\hat{N}}\cdots \Tr_1(\sqrt{\hat{N}}}_{k\,\mathrm{times}}
			\sqrt{\eta}^{\hat{N}}\hat{\rho}
			\sqrt{\eta}^{\hat{N}} \underbrace{\sqrt{\hat{N}})\cdots\sqrt{\hat{N}})}_{k\,\mathrm{times}}
			.
			\label{eq:BS loss in terms of us}
	\end{aligned}\end{equation} The first transformation of Eq.~\eqref{eq:BS loss in terms of us} matches the one found in Refs. \cite{Oszmaniecetal2016,OszmaniecBrod2018} in the contexts of boson sampling and quantum metrology.  This provides an alternate formula for loss channels and directly explains why equal loss on $d$ modes commutes with all linear optical networks that enact $\ha_i\to\sum_j U_{ij}\ha_j$ (because the $\Tr_1$ and $\hat{N}$ operations commute with such networks).
	
	\section{Example applications}
	\subsection{Purity of reduced state}
	One indicator of quantum polarization properties is how mixed a pure state ${\hat{\rho}}_N$ is after removing a certain number of photons \cite{Rudzinskietal2023arxiv}. With our formula from Eq.~\eqref{eq:Tr1 general}, we can express the purity of a reduced pure state as
	\begin{equation}\begin{aligned}
			\Tr\{[\Tr_{N-q}(\ket{\psi_N}\bra{\psi_N})]^2\}&=\frac{q!^2}{N!^2}\sum_{i_1\cdots i_{N-q},j_1\cdots j_{N-q}}|\langle \ha^\dagger_{i_1}\cdots \ha^\dagger_{i_{N-q}} \ha_{j_1}\cdots\ha_{j_{N-q}}\rangle|^2\\
			&=\frac{q!^2}{N!^2}\sum_{\mathbf{k}\mathbf{l}}\binom{N-q}{\mathbf{k}}\binom{N-q}{\mathbf{l}}|\langle \hat{O}_{\mathbf{k}\mathbf{l}}\rangle|^2.
	\end{aligned}\end{equation} This is just a sum of the $N-q$th-order moments of the state, as the multinomial coefficients vanish unless $|\mathbf{k}|=|\mathbf{l}|=N-q$. For example, in the two-mode case of polarization, the purity of a state after tracing out a single photon is determined by the Stokes parameters because it exclusively depends on $\langle \ha^\dagger_i\ha_j\rangle$; the purity after tracing out two photons depends on the covariances of the Stokes parameters; etc. We can explicitly calculate for the Stokes parameters that the purity of the state after tracing out a single photon is proportional to $S_0^2+S_1^2+S_2^2+S_3^2$, such that it is maximal for spin-coherent states and minimal for pure states whose Stokes vector vanishes.
	
	\subsection{Photon-number projection}
	Our $q$-photon state ${\hat{\varrho}}(q|{\hat{\rho}})$ from Eq.~\eqref{eq:rho random q} is not the same as the projection of a state ${\hat{\rho}}$ onto the $q$-photon subspace; randomly choosing $q$ photons is different from projecting onto $q$ photons. How do these two ideas compare?
	
	Due to the relation $\ket{0}\bra{0}=:\exp(-\had\ha):$ that uses the normal ordering operation $::$, we can express a projector onto a certain $m$-photon state as
	\begin{equation}
		\ket{m}\bra{m}=\frac{1}{m!}\ha^{\dagger m}\sum_{n=0}^\infty (-1)^n\frac{\ha^{\dagger n}\ha^n}{n!}\ha^m=\sum_{n=m}^\infty\frac{(-1)^{n-m}}{m!(n-m)!}\hat{O}_{nn}.
	\end{equation} Evaluating the expectation value of the projector for a state will thus require all states ${\hat{\varrho}}(q|{\hat{\rho}})$ with $q\geq m$. Intuitively, this means that there is a contribution to the projection from all components of the state from which $m$ photons \textit{could have} arisen: we start with a contribution from the $m$-photon random state, adjust for contributions that may have arisen from states that truly had $m+1$ photons out of which $m$ were randomly chosen by considering the $m+1$-photon random state, and so on. 
	
	This result holds for multimode states as well. To project onto the $m$-photon subspace of a multimode state, one needs to account for all of the components of the state from which $m$-photons may be randomly selected using our above prescription.
	
	\section{Conclusion}
	
	There is a unique unambiguous method for describing how a beam of light behaves when $k$ photons are removed from it or when $q$ photons are selected from it at random. We have provided a number of simple formulae for these computations, with the result that the $q$th-order correlation functions of a $d$-mode state are in one-to-one correspondence with the expansion coefficients in the Fock basis for $q$ photons selected at random from the state. This closes the loop between the mathematics and intuition of why singles counts, doubles counts, and more are required for observing particular correlations; a detector effectively sees a $q$-photon state chosen at random from the original state. All of our work has been derived in the language of quantum optics, but the results should find equal application in the study of any totally symmetric $d$-level system across quantum information theory.

	\begin{backmatter}
		\bmsection{Funding}
		NSERC PDF program.
		
		\bmsection{Acknowledgments}
		The NRC headquarters is located on the traditional unceded territory of the Algonquin Anishinaabe and Mohawk peoples. 
		
		\bmsection{Disclosures}
		The authors declare no conflicts of interest.

		\bmsection{Data Availability Statement}
		No data were generated or analyzed in the presented research.
		
		\bigskip

		\bmsection{Supplemental document}
		See Supplements 1 and 2 for supporting content. 
		
	\end{backmatter}


\begin{thebibliography}{10}
		\newcommand{\enquote}[1]{``#1''}
		
		\bibitem{Glauber1963}
		R.~J. Glauber, \enquote{Coherent and incoherent states of the radiation field,}
		{\protect\JournalTitle{Physical Review}} \textbf{131}, 2766--2788 (1963).
		
		\bibitem{Glauber1963quantumtheorycoherence}
		R.~J. Glauber, \enquote{The quantum theory of optical coherence,}
		{\protect\JournalTitle{Physical Review}} \textbf{130}, 2529--2539 (1963).
		
		\bibitem{Glauber1963correlations}
		R.~J. Glauber, \enquote{Photon correlations,} {\protect\JournalTitle{Phys. Rev.
				Lett.}} \textbf{10}, 84--86 (1963).
		
		\bibitem{Sudarshan1963}
		E.~C.~G. Sudarshan, \enquote{Equivalence of semiclassical and quantum
			mechanical descriptions of statistical light beams,}
		{\protect\JournalTitle{Physical Review Letters}} \textbf{10}, 277--279
		(1963).
		
		\bibitem{MehtaSudarshan1965}
		C.~L. Mehta and E.~C.~G. Sudarshan, \enquote{Relation between quantum and
			semiclassical description of optical coherence,} {\protect\JournalTitle{Phys.
				Rev.}} \textbf{138}, B274--B280 (1965).
		
		\bibitem{TitulaerGlauber1965}
		U.~M. Titulaer and R.~J. Glauber, \enquote{Correlation functions for coherent
			fields,} {\protect\JournalTitle{Phys. Rev.}} \textbf{140}, B676--B682 (1965).
		
		\bibitem{CahillGlauber1969}
		K.~E. Cahill and R.~J. Glauber, \enquote{Ordered expansions in boson amplitude
			operators,} {\protect\JournalTitle{Phys. Rev.}} \textbf{177}, 1857--1881
		(1969).
		
		\bibitem{AgarwalWolf1970}
		G.~S. Agarwal and E.~Wolf, \enquote{Calculus for functions of noncommuting
			operators and general phase-space methods in quantum mechanics. i. mapping
			theorems and ordering of functions of noncommuting operators,}
		{\protect\JournalTitle{Phys. Rev. D}} \textbf{2}, 2161--2186 (1970).
		
		\bibitem{CarmichaelWalls1976}
		H.~J. Carmichael and D.~F. Walls, \enquote{Proposal for the measurement of the
			resonant stark effect by photon correlation techniques,}
		{\protect\JournalTitle{Journal of Physics B: Atomic and Molecular Physics}}
		\textbf{9}, L43 (1976).
		
		\bibitem{Loudon1976}
		R.~Loudon, \enquote{Photon bunching and antibunching,}
		{\protect\JournalTitle{Physics Bulletin}} \textbf{27}, 21 (1976).
		
		\bibitem{Kimbleetal1977}
		H.~J. Kimble, M.~Dagenais, and L.~Mandel, \enquote{Photon antibunching in
			resonance fluorescence,} {\protect\JournalTitle{Phys. Rev. Lett.}}
		\textbf{39}, 691--695 (1977).
		
		\bibitem{Leuchs1986}
		G.~Leuchs, \emph{Photon Statistics, Antibunching and Squeezed States} (Springer
		US, Boston, MA, 1986), pp. 329--360.
		
		\bibitem{Grangieretal1986}
		P.~Grangier, G.~Roger, and A.~Aspect, \enquote{Experimental evidence for a
			photon anticorrelation effect on a beam splitter: A new light on
			single-photon interferences,} {\protect\JournalTitle{Europhysics Letters}}
		\textbf{1}, 173 (1986).
		
		\bibitem{Rempeetal1990}
		G.~Rempe, F.~Schmidt-Kaler, and H.~Walther, \enquote{Observation of
			sub-poissonian photon statistics in a micromaser,}
		{\protect\JournalTitle{Phys. Rev. Lett.}} \textbf{64}, 2783--2786 (1990).
		
		\bibitem{Mandel1979}
		L.~Mandel, \enquote{Sub-poissonian photon statistics in resonance
			fluorescence,} {\protect\JournalTitle{Opt. Lett.}} \textbf{4}, 205--207
		(1979).
		
		\bibitem{Mandel1986}
		L.~Mandel, \enquote{Non-classical states of the electromagnetic field,}
		{\protect\JournalTitle{Physica Scripta}} \textbf{1986}, 34 (1986).
		
		\bibitem{AgarwalTara1991}
		G.~S. Agarwal and K.~Tara, \enquote{Nonclassical properties of states generated
			by the excitations on a coherent state,} {\protect\JournalTitle{Phys. Rev.
				A}} \textbf{43}, 492--497 (1991).
		
		\bibitem{Klyshko1996}
		D.~Klyshko, \enquote{Observable signs of nonclassical light,}
		{\protect\JournalTitle{Physics Letters A}} \textbf{213}, 7--15 (1996).
		
		\bibitem{Lee1998}
		C.~T. Lee, \enquote{Simple criterion for nonclassical two-mode states,}
		{\protect\JournalTitle{J. Opt. Soc. Am. B}} \textbf{15}, 1187--1191 (1998).
		
		\bibitem{RichterVogel2002}
		T.~Richter and W.~Vogel, \enquote{Nonclassicality of quantum states: A
			hierarchy of observable conditions,} {\protect\JournalTitle{Phys. Rev.
				Lett.}} \textbf{89}, 283601 (2002).
		
		\bibitem{ShchukinVogel2005}
		E.~V. Shchukin and W.~Vogel, \enquote{Nonclassical moments and their
			measurement,} {\protect\JournalTitle{Phys. Rev. A}} \textbf{72}, 043808
		(2005).
		
		\bibitem{Achillesetal2006}
		D.~Achilles, C.~Silberhorn, and I.~A. Walmsley, \enquote{Direct, loss-tolerant
			characterization of nonclassical photon statistics,}
		{\protect\JournalTitle{Phys. Rev. Lett.}} \textbf{97}, 043602 (2006).
		
		\bibitem{Zavattaetal2007}
		A.~Zavatta, V.~Parigi, and M.~Bellini, \enquote{Experimental nonclassicality of
			single-photon-added thermal light states,} {\protect\JournalTitle{Phys. Rev.
				A}} \textbf{75}, 052106 (2007).
		
		\bibitem{Miranowiczetal2010}
		A.~Miranowicz, M.~Bartkowiak, X.~Wang, Y.-x. Liu, and F.~Nori, \enquote{Testing
			nonclassicality in multimode fields: A unified derivation of classical
			inequalities,} {\protect\JournalTitle{Phys. Rev. A}} \textbf{82}, 013824
		(2010).
		
		\bibitem{Ryletal2015}
		S.~Ryl, J.~Sperling, E.~Agudelo, M.~Mraz, S.~K\"ohnke, B.~Hage, and W.~Vogel,
		\enquote{Unified nonclassicality criteria,} {\protect\JournalTitle{Phys. Rev.
				A}} \textbf{92}, 011801 (2015).
		
		\bibitem{SahotaQuesada2015}
		J.~Sahota and N.~Quesada, \enquote{Quantum correlations in optical metrology:
			Heisenberg-limited phase estimation without mode entanglement,}
		{\protect\JournalTitle{Physical Review A}} \textbf{91}, 013808 (2015).
		
		\bibitem{Hartmannetal2015}
		S.~Hartmann, F.~Friedrich, A.~Molitor, M.~Reichert, W.~Elsäßer, and
		R.~Walser, \enquote{Tailored quantum statistics from broadband states of
			light,} {\protect\JournalTitle{New Journal of Physics}} \textbf{17}, 043039
		(2015).
		
		\bibitem{Perinaetal2017}
		J.~Pe\ifmmode~\check{r}\else \v{r}\fi{}ina, V.~Mich\'alek, and O.~c.~v.
		Haderka, \enquote{Higher-order sub-poissonian-like nonclassical fields:
			Theoretical and experimental comparison,} {\protect\JournalTitle{Phys. Rev.
				A}} \textbf{96}, 033852 (2017).
		
		\bibitem{Sudhiretal2017}
		V.~Sudhir, R.~Schilling, S.~A. Fedorov, H.~Sch\"utz, D.~J. Wilson, and T.~J.
		Kippenberg, \enquote{Quantum correlations of light from a room-temperature
			mechanical oscillator,} {\protect\JournalTitle{Phys. Rev. X}} \textbf{7},
		031055 (2017).
		
		\bibitem{KorolkovaLeuchs2019}
		N.~Korolkova and G.~Leuchs, \enquote{Quantum correlations in separable
			multi-mode states and in classically entangled light,}
		{\protect\JournalTitle{Reports on Progress in Physics}} \textbf{82}, 056001
		(2019).
		
		\bibitem{Malpanietal2019}
		P.~Malpani, N.~Alam, K.~Thapliyal, A.~Pathak, V.~Narayanan, and S.~Banerjee,
		\enquote{Lower- and higher-order nonclassical properties of photon added and
			subtracted displaced fock states,} {\protect\JournalTitle{Annalen der
				Physik}} \textbf{531}, 1800318 (2019).
		
		\bibitem{Madsenetal2022}
		L.~S. Madsen, F.~Laudenbach, M.~F. Askarani, F.~Rortais, T.~Vincent, J.~F.~F.
		Bulmer, F.~M. Miatto, L.~Neuhaus, L.~G. Helt, M.~J. Collins, A.~E. Lita,
		T.~Gerrits, S.~W. Nam, V.~D. Vaidya, M.~Menotti, I.~Dhand, Z.~Vernon,
		N.~Quesada, and J.~Lavoie, \enquote{Quantum computational advantage with a
			programmable photonic processor,} {\protect\JournalTitle{Nature}}
		\textbf{606}, 75--81 (2022).
		
		\bibitem{Goldberg2022}
		A.~Z. Goldberg, \enquote{Chapter three - quantum polarimetry,}  (Elsevier,
		2022), pp. 185--274.
		
		\bibitem{Klyshko1997}
		D.~M. Klyshko, \enquote{Polarization of light: Fourth-order effects and
			polarization-squeezed states,} {\protect\JournalTitle{Journal of Experimental
				and Theoretical Physics}} \textbf{84}, 1065--1079 (1997).
		
		\bibitem{Luis2016}
		A.~Luis, \enquote{Polarization in quantum optics,}
		{\protect\JournalTitle{Progress in Optics}} \textbf{61}, 283 -- 331 (2016).
		
		\bibitem{Goldbergetal2021polarization}
		A.~Z. Goldberg, P.~de~la Hoz, G.~Bj\"{o}rk, A.~B. Klimov, M.~Grassl, G.~Leuchs,
		and L.~L. S\'{a}nchez-Soto, \enquote{Quantum concepts in optical
			polarization,} {\protect\JournalTitle{Advances in Optics and Photonics}}
		\textbf{13}, 1--73 (2021).
		
		\bibitem{Young1801}
		T.~Young, \enquote{I. the bakerian lecture. experiments and calculations
			relative to physical optics,} {\protect\JournalTitle{Philosophical
				Transactions of the Royal Society of London}} \textbf{94}, 1--16 (1804).
		
		\bibitem{ArnaudCerf2013}
		L.~Arnaud and N.~J. Cerf, \enquote{Exploring pure quantum states with maximally
			mixed reductions,} {\protect\JournalTitle{Phys. Rev. A}} \textbf{87}, 012319
		(2013).
		
		\bibitem{Baguetteetal2014}
		D.~Baguette, T.~Bastin, and J.~Martin, \enquote{Multiqubit symmetric states
			with maximally mixed one-qubit reductions,} {\protect\JournalTitle{Phys. Rev.
				A}} \textbf{90}, 032314 (2014).
		
		\bibitem{Giraudetal2015}
		O.~Giraud, D.~Braun, D.~Baguette, T.~Bastin, and J.~Martin, \enquote{Tensor
			representation of spin states,} {\protect\JournalTitle{Physical Review
				Letters}} \textbf{114}, 080401 (2015).
		
		\bibitem{BaguetteMartin2017}
		D.~Baguette and J.~Martin, \enquote{Anticoherence measures for pure spin
			states,} {\protect\JournalTitle{Physical Review A}} \textbf{96}, 032304
		(2017).
		
		\bibitem{AaronsonBrod2016}
		S.~Aaronson and D.~J. Brod, \enquote{Bosonsampling with lost photons,}
		{\protect\JournalTitle{Phys. Rev. A}} \textbf{93}, 012335 (2016).
		
		\bibitem{Uedaetal1990}
		M.~Ueda, N.~Imoto, and T.~Ogawa, \enquote{Quantum theory for continuous
			photodetection processes,} {\protect\JournalTitle{Phys. Rev. A}} \textbf{41},
		3891--3904 (1990).
		
		\bibitem{Daknaetal1997}
		M.~Dakna, T.~Anhut, T.~Opatrn\'y, L.~Kn\"oll, and D.-G. Welsch,
		\enquote{Generating schr\"odinger-cat-like states by means of conditional
			measurements on a beam splitter,} {\protect\JournalTitle{Phys. Rev. A}}
		\textbf{55}, 3184--3194 (1997).
		
		\bibitem{Opatrnyetal2000}
		T.~Opatrn\'y, G.~Kurizki, and D.-G. Welsch, \enquote{Improvement on
			teleportation of continuous variables by photon subtraction via conditional
			measurement,} {\protect\JournalTitle{Phys. Rev. A}} \textbf{61}, 032302
		(2000).
		
		\bibitem{MizrahiDodonov2002}
		S.~S. Mizrahi and V.~V. Dodonov, \enquote{Creating quanta with an
			‘annihilation’ operator,} {\protect\JournalTitle{Journal of Physics A:
				Mathematical and General}} \textbf{35}, 8847 (2002).
		
		\bibitem{Wengeretal2004}
		J.~Wenger, R.~Tualle-Brouri, and P.~Grangier, \enquote{Non-gaussian statistics
			from individual pulses of squeezed light,} {\protect\JournalTitle{Phys. Rev.
				Lett.}} \textbf{92}, 153601 (2004).
		
		\bibitem{Ourjoumtsevetal2006}
		A.~Ourjoumtsev, R.~Tualle-Brouri, J.~Laurat, and P.~Grangier,
		\enquote{Generating optical schrödinger kittens for quantum information
			processing,} {\protect\JournalTitle{Science}} \textbf{312}, 83--86 (2006).
		
		\bibitem{Parigietal2007}
		V.~Parigi, A.~Zavatta, M.~Kim, and M.~Bellini, \enquote{Probing quantum
			commutation rules by addition and subtraction of single photons to/from a
			light field,} {\protect\JournalTitle{Science}} \textbf{317}, 1890--1893
		(2007).
		
		\bibitem{Zavattaetal2008}
		A.~Zavatta, V.~Parigi, M.~S. Kim, and M.~Bellini, \enquote{Subtracting photons
			from arbitrary light fields: experimental test of coherent state invariance
			by single-photon annihilation,} {\protect\JournalTitle{New Journal of
				Physics}} \textbf{10}, 123006 (2008).
		
		\bibitem{Takahashietal2008}
		H.~Takahashi, K.~Wakui, S.~Suzuki, M.~Takeoka, K.~Hayasaka, A.~Furusawa, and
		M.~Sasaki, \enquote{Generation of large-amplitude coherent-state
			superposition via ancilla-assisted photon subtraction,}
		{\protect\JournalTitle{Phys. Rev. Lett.}} \textbf{101}, 233605 (2008).
		
		\bibitem{Gerritsetal2010}
		T.~Gerrits, S.~Glancy, T.~S. Clement, B.~Calkins, A.~E. Lita, A.~J. Miller,
		A.~L. Migdall, S.~W. Nam, R.~P. Mirin, and E.~Knill, \enquote{Generation of
			optical coherent-state superpositions by number-resolved photon subtraction
			from the squeezed vacuum,} {\protect\JournalTitle{Phys. Rev. A}} \textbf{82},
		031802 (2010).
		
		\bibitem{NavarreteBenlloch2012}
		C.~Navarrete-Benlloch, R.~Garc\'{\i}a-Patr\'on, J.~H. Shapiro, and N.~J. Cerf,
		\enquote{Enhancing quantum entanglement by photon addition and subtraction,}
		{\protect\JournalTitle{Phys. Rev. A}} \textbf{86}, 012328 (2012).
		
		\bibitem{Bartleyetal2013sub}
		T.~J. Bartley, P.~J.~D. Crowley, A.~Datta, J.~Nunn, L.~Zhang, and I.~Walmsley,
		\enquote{Strategies for enhancing quantum entanglement by local photon
			subtraction,} {\protect\JournalTitle{Phys. Rev. A}} \textbf{87}, 022313
		(2013).
		
		\bibitem{FanZubairy2018}
		L.~Fan and M.~S. Zubairy, \enquote{Quantum illumination using non-gaussian
			states generated by photon subtraction and photon addition,}
		{\protect\JournalTitle{Phys. Rev. A}} \textbf{98}, 012319 (2018).
		
		\bibitem{Stiesdaletal2021}
		N.~Stiesdal, H.~Busche, K.~Kleinbeck, J.~Kumlin, M.~G.~Hansen, H.~P. Büchler,
		and S.~Hofferberth, \enquote{Controlled multi-photon subtraction with
			cascaded Rydberg superatoms as single-photon absorbers,}
		{\protect\JournalTitle{Nature Communications}} \textbf{12}, 4328 (2021).
		
		\bibitem{Takaseetal2021}
		K.~Takase, J.-i. Yoshikawa, W.~Asavanant, M.~Endo, and A.~Furusawa,
		\enquote{Generation of optical schr\"odinger cat states by generalized photon
			subtraction,} {\protect\JournalTitle{Phys. Rev. A}} \textbf{103}, 013710
		(2021).
		
		\bibitem{Melalkiaetal2022}
		M.~F. Melalkia, L.~Brunel, S.~Tanzilli, J.~Etesse, and V.~D'Auria,
		\enquote{Theoretical framework for photon subtraction with
			non--mode-selective resources,} {\protect\JournalTitle{Phys. Rev. A}}
		\textbf{105}, 013720 (2022).
		
		\bibitem{Nunnetal2022}
		C.~M. Nunn, J.~D. Franson, and T.~B. Pittman, \enquote{Modifying quantum
			optical states by zero-photon subtraction,} {\protect\JournalTitle{Phys. Rev.
				A}} \textbf{105}, 033702 (2022).
		
		\bibitem{Nevenetal2018}
		A.~Neven, J.~Martin, and T.~Bastin, \enquote{Entanglement robustness against
			particle loss in multiqubit systems,} {\protect\JournalTitle{Phys. Rev. A}}
		\textbf{98}, 062335 (2018).
		
		\bibitem{Quintaetal2019}
		G.~M. Quinta, R.~Andr\'e, A.~Burchardt, and K.~\ifmmode~\dot{Z}\else
		\.{Z}\fi{}yczkowski, \enquote{Cut-resistant links and multipartite
			entanglement resistant to particle loss,} {\protect\JournalTitle{Phys. Rev.
				A}} \textbf{100}, 062329 (2019).
		
		\bibitem{OszmaniecBrod2018}
		M.~Oszmaniec and D.~J. Brod, \enquote{Classical simulation of photonic linear
			optics with lost particles,} {\protect\JournalTitle{New Journal of Physics}}
		\textbf{20}, 092002 (2018).
		
		\bibitem{Mikhailov1983}
		V.~V. Mikhailov, \enquote{Ordering of some boson operator functions,}
		{\protect\JournalTitle{Journal of Physics A: Mathematical and General}}
		\textbf{16}, 3817 (1983).
		
		\bibitem{Blasiaketal2003}
		P.~Blasiak, K.~Penson, and A.~Solomon, \enquote{The general boson normal
			ordering problem,} {\protect\JournalTitle{Physics Letters A}} \textbf{309},
		198--205 (2003).
		
		\bibitem{Oszmaniecetal2016}
		M.~Oszmaniec, R.~Augusiak, C.~Gogolin, J.~Ko\l{}ody\ifmmode~\acute{n}\else
		\'{n}\fi{}ski, A.~Ac\'{\i}n, and M.~Lewenstein, \enquote{Random bosonic
			states for robust quantum metrology,} {\protect\JournalTitle{Phys. Rev. X}}
		\textbf{6}, 041044 (2016).
		
		\bibitem{Rudzinskietal2023arxiv}
		M.~Rudziński, A.~Burchardt, and K.~Życzkowski, \enquote{Orthonormal bases of
			extreme spin coherence,} 	arXiv:2306.00532 (2023).
		
	\end{thebibliography}
\end{document}


\title{
Supplement for ``Correlations
for symmetric states: what subsets of photons are doing within a beam of light''}

\author{Aaron Z. Goldberg,\authormark{1,2,*}}

\address{\authormark{1}National Research Council of Canada, 100 Sussex Drive, Ottawa, Ontario K1N 5A2, Canada\\
\authormark{2}Department of Physics, University of Ottawa, Advanced Research Complex, 25 Templeton Street, Ottawa, Ontario K1N 6N5, Canada}

\email{\authormark{*}aaron.goldberg@nrc-cnrc.gc.ca} 


\section{Supplement 1}
We here directly prove that
\eq{
{\hat{\varrho}}(q|\hat{\rho})\equiv \sum_N P(q\,\mathrm{from}\,N) {\hat{\varrho}}(q|\hat{\rho}_N) \propto \sum_{|\mathbf{m}|=|\mathbf{n}|=q}\frac{\langle \hat{O}_{\mathbf{m}\mathbf{n}}\rangle_{\hat{\rho}}}{\sqrt{m_1!\cdots m_d!n_1!\cdots n_d!}}\ket{\mathbf{n}}\bra{\mathbf{m}}.
}

We begin with our arbitrary state expressed in the Fock basis as $\hat{\rho}=\sum_{\mathbf{k}\mathbf{j}}\hat{\rho}_{\mathbf{k}\mathbf{j}}\ket{\mathbf{k}}\bra{\mathbf{j}}$. Using the projectors onto the $N$-photon subspaces,
\eq{
\hat{P}_N=\sum_{|\mathbf{k}|=N}\ket{\mathbf{k}}\bra{\mathbf{k}},
} we find the $N$-photon components of the state to be
\eq{
p_N\hat{\rho}_N\equiv \hat{P}_N\hat{\rho}\hat{P}_N=\sum_{|\mathbf{k}|=|\mathbf{j}|=N}\hat{\rho}_{\mathbf{k}\mathbf{j}}\ket{\mathbf{k}}\bra{\mathbf{j}}.
} Note that we set the states $\hat{\rho}_N$ to be normalized to unity, with $p_N$ the corresponding probability of the state actually having $N$ photons. To trace out $N-q$ photons, we apply \eq{\Tr_{N- q}(p_N\hat{\rho}_N) &=p_N\frac{q!}{N!}\sum_{i_1,\cdots,i_{N-q}}\ha_{i_1}\cdots \ha_{i_{N-q}}\hat{\rho}_N\ha^\dagger_{i_{N-q}}\cdots\ha^\dagger_{i_1}
\\
&=p_N\frac{q!}{N!}\sum_{|\mathbf{l}|=N-q}\binom{N-q}{\mathbf{l}}\ha_{1}^{l_1}\cdots \ha_{d}^{l_d}\hat{\rho}_N\ha^{\dagger l_1}_{1}\cdots \ha_{d}^{\dagger l_d},
}where we use the multinomial coefficient because the creation operators for each mode commute with each other. Now the action on a term $\ket{\mathbf{k}}\bra{\mathbf{j}}$ requires $k_i-l_i\geq 0$ and $j_i-l_i\geq 0$ for all modes $i$. We find the above to equal
\eq{
\mathrm{Tr}_{N-q}(p_N\hat{\rho}_N)&=
\frac{q!}{N!}\sum_{|\mathbf{l}|=N-q}\sum_{|\mathbf{k}|=|\mathbf{j}|=N}\binom{N-q}{\mathbf{l}}\sqrt{\frac{k_1!\cdots k_d! j_1!\cdots j_d!}{(k_1-l_1)!\cdots (k_d-l_d)!(j_1-l_1)!\cdots (j_d-l_d)!}}\\
&\quad\times\hat{\rho}_{\mathbf{k}\mathbf{j}}\ket{\mathbf{k}-\mathbf{l}}\bra{\mathbf{j}-\mathbf{l}}\\
&=
\binom{N}{q}^{-1}\sum_{|\mathbf{m}|=|\mathbf{n}|=q}\sum_{|\mathbf{k}|=|\mathbf{j}|=N}\sqrt{\binom{k_1}{m_1}\cdots \binom{k_d}{m_d}\binom{j_1}{n_1}\cdots \binom{j_d}{n_d}}\hat{\rho}_{\mathbf{k},\mathbf{j}}\ket{\mathbf{m}}\bra{\mathbf{n}}\delta_{\mathbf{k}-\mathbf{m},\mathbf{j}-\mathbf{n}}.
} The binomial coefficients restrict the sums to the appropriate domains.

Next, we identify $P(q\,\mathrm{from}\,N)=p_N\binom{N}{q}/\mathcal{N}(q)$ as the probability of $q$ photons originating from the $N$-photon sector and ${\hat{\varrho}}(q|\hat{\rho}_N)=\Tr_{N-q}(\hat{\rho}_N)$ as the state from which they must have originated. Then we can combine our calculations for the entire state to yield
\eq{
{\hat{\varrho}}(q|\hat{\rho})&=\frac{1}{\mathcal{N}(q)}\sum_N\sum_{|\mathbf{m}|=|\mathbf{n}|=q}\sum_{|\mathbf{k}|=|\mathbf{j}|=N}\sqrt{\binom{k_1}{m_1}\cdots \binom{k_d}{m_d}\binom{j_1}{n_1}\cdots \binom{j_d}{n_d}}\hat{\rho}_{\mathbf{k},\mathbf{j}}\ket{\mathbf{m}}\bra{\mathbf{n}}\delta_{\mathbf{k}-\mathbf{m},\mathbf{j}-\mathbf{n}}\\
&=\frac{1}{\mathcal{N}(q)}\sum_{|\mathbf{m}|=|\mathbf{n}|=q}\frac{\langle\hat{O}_{\mathbf{n}\mathbf{m}}\rangle_{\hat{\rho}}}{\sqrt{m_1!\cdots m_d!n_1!\cdots n_d!}}\ket{\mathbf{m}}\bra{\mathbf{n}}.
} Swapping the index labels $\mathbf{m}\leftrightarrow\mathbf{n}$ reproduces the result in the main text.

\section{Supplement 2}
We next explain the connection between loss channels as usually discussed in the literature and our formulation. Typically, a loss channel for mode $i$ is described by enacting the input-output relation
\eq{
\ha_i\to\sqrt{\eta_i}\ha_i+\sqrt{1-\eta_i}\hb_i
} and then tracing out the bosonic mode annihilated by $\hb_i$ that began in its vacuum state. One can notice by these input-output relations that, if every channel has the same transmission parameter $\eta_i=\eta$, then the a linear optical transformation $\ha_i\to\sum_{ij}U_j \ha_j$ commutes with the overall loss channels because their actions simply differ by a unitary transformation among the vacuum modes. It is striking to compare this with our operation $\Tr_1$, which also commutes with the action of linear optical networks and thus has the same effect whether it acts before or after such a network. Surely these two phenomena must be connected, and indeed they are.

A loss channel enacts a beam splitter between modes $\ha_i$ and $\hb_i$. Its action is equivalent to acting on a global state supplanted with an auxiliary vacuum mode that gets traced out:
\eq{
\hat{\rho}\underset{\eta_i}{\to} \sum_{n}\vphantom{a}_{b_i}\langle n|\hat{B}(\eta_i) (\hat{\rho}\otimes\ket{0}_{b_i}\bra{0})\hat{B}^\dagger(\eta_i) \ket{n}_{b_i}.
} The beam-splitter operation can take many equivalent forms, the most useful of which for our purposes is
\eq{
\hat{B}(\eta_i)=&\exp(\sqrt{\frac{1-\eta_i}{\eta_i}}\ha_i\hb^\dagger_i)\exp((\ha_i^\dagger\ha_i-\hb_i^\dagger\hb_i)\ln\sqrt{\eta_i})\exp(-\sqrt{\frac{1-\eta_i}{\eta_i}}\ha_i^\dagger\hb_i)\\
\Rightarrow \quad\vphantom{a}_{b_i}\bra{n}\hat{B}(\eta_i)\ket{0}_{b_i}=&\vphantom{a}_{b_i}\bra{n}\exp(\sqrt{\frac{1-\eta_i}{\eta_i}}\ha_i\hb^\dagger_i)\ket{0}_{b_i}\exp(\ha_i^\dagger\ha_i\ln\sqrt{\eta_i})\\
=&\frac{1}{\sqrt{n!}}\left(\frac{1-\eta_i}{\eta_i}\right)^{n/2}\ha_i^n \sqrt{\eta_i}^{\ha_i^\dagger\ha_i},
} which can be verified for its action $\hat{B}(\eta_i)
\ha_i\hat{B}^\dagger(\eta_i)=\sqrt{\eta_i}\ha_i+\sqrt{1-\eta_i}\hb_i$. When each mode has the same loss channel, the state evolves as
\eq{
\hat{\rho}\underset{\eta}{\to}&\sum_{n_1,\cdots,n_d}\frac{1}{n_1!\cdots n_d!}\left(\frac{1-\eta}{\eta}\right)^{n_1+\cdots+n_d}\ha_1^{n_1}\sqrt{\eta}^{\ha_1^\dagger \ha_1}\cdots \ha_d^{n_d}\sqrt{\eta}^{\ha_d^\dagger \ha_d}\hat{\rho}
\sqrt{\eta}^{\ha_d^\dagger \ha_d}\ha_d^{\dagger n_d}\cdots \sqrt{\eta}^{\ha_1^\dagger \ha_1}\ha_1^{\dagger n_1}\\
&=\sum_N\frac{\left(\frac{1-\eta}{\eta}\right)^{N}}{N!}\sum_{|\mathbf{n}|=N}\binom{N}{\mathbf{n}}\ha_1^{n_1}\cdots \ha_d^{n_d}\sqrt{\eta}^{\hat{N}}\hat{\rho}
\sqrt{\eta}^{\hat{N}}\ha_d^{\dagger n_d}\cdots \ha_1^{\dagger n_1}.
} The multinomial coefficient tells us how many times each distribution of modes operators should be counted; this is equivalent to having $N$ different operators that can each come from one of $d$ different modes. By using the alternate counting, we find
\eq{
\hat{\rho}\underset{\eta}{\to}&
\sum_N\frac{\left(\frac{1-\eta}{\eta}\right)^{N}}{N!}\sum_{i_1=1}^d\cdots \sum_{i_N=1}^d \ha_{i_1}\cdots \ha_{i_N}\sqrt{\eta}^{\hat{N}}\hat{\rho}
\sqrt{\eta}^{\hat{N}}\ha_{i_N}^{\dagger}\cdots \ha_{i_1}^{\dagger}\\
&=\sum_N\frac{\left(\frac{1-\eta}{\eta}\right)^{N}}{N!} 
\underbrace{\Tr_1(\sqrt{\hat{N}}\cdots }_{N-1\,\mathrm{more\, times}}
\Tr_1(\sqrt{\hat{N}}\sqrt{\eta}^{\hat{N}}\hat{\rho}
\sqrt{\eta}^{\hat{N}} \sqrt{\hat{N}})\cdots\sqrt{\hat{N}}).
} The continuous loss channel can thus be expressed in terms of convex combinations tracing out individual photons $N$ times, but with a photon-number-dependent factor being applied before each individual photon is traced out. Each component of this expression is independent from mode decomposition and thus commutes with the action of linear optical transformations on $\hat{\rho}$.

\begin{backmatter}
\bmsection{Funding}
NSERC PDF program.

\bmsection{Acknowledgments}
The NRC headquarters is located on the traditional unceded territory of the Algonquin Anishinaabe and Mohawk peoples. 

\bmsection{Disclosures}
The authors declare no conflicts of interest.

\bmsection{Data Availability Statement}
No data were generated or analyzed in the presented research.

\bigskip

\end{backmatter}
